\documentclass[twocolumn]{elsart}
\usepackage{natbib}
\usepackage{graphicx}
\usepackage{aas_macros}

\begin{document}
\begin{frontmatter}
\title{Ultra-compact (X-ray) binaries}
\author{G. Nelemans}
\address{Department of Astrophysics, IMAPP, Radboud University Nijmegen, NL}
\author{P.G. Jonker}
\address{SRON National Institute for Space Research, Utrecht, NL \\
Harvard--Smithsonian Center for Astrophysics, Cambridge, USA}

\begin{abstract}
A short review of ultra-compact binaries, focused on ultra-compact
X-ray binaries, is followed by a discussion of recent results of our
VLT campaign to obtain optical spectra of (candidate) ultra-compact
X-ray binaries. We find evidence for carbon/oxygen as well as
helium/nitrogen discs and no evidence for (traces) of hydrogen. This
suggests that the donors in the observed systems are white
dwarfs. However, we also find large differences between the two C/O
discs of which we have good spectra, which highlights the need for a
better understanding of the optical spectra.
\end{abstract}
\end{frontmatter}

\section{Introduction: ultra-compact binaries}

Ultra-compact binaries are double stars with periods less than about
one hour. In such short-period binaries the components must be so
close together that ordinary, hydrogen rich, stars do not fit in
\citep[e.g.][]{nrj86}. Both stars in ultra-compact binaries thus must
be compact stars: the cores of evolved giants, white dwarfs, neutron
stars or black holes.

Of special interest are ultra-compact binaries in which mass is
transferred from one component to the other. Observationally two types
are known: ones with an white dwarf accretor (called AM CVn systems,
after the first detected system) and ones with a neutron star
accretor, called ultra-compact X-ray binaries (UCXBs).  There are
currently 16 AM CVn stars known, plus two candidate systems. For more
information we refer to a recent review \citep{nel05}. Four new
systems were reported in the last year, found in the SDSS
spectroscopic database \citep{ahh+05}. The number of known systems is
increasing rapidly: 9 of the known systems and one of the two
candidates were discovered in the last five years. The two candidate
systems V407 Vul and RX J0806.3+1527, are both found as X-ray sources,
showing a 100 per cent modulation on 9.5 and 5.4 mins
respectively. One interpretation is that these periods are orbital
periods and thus they could be part of the AM CVn family. However,
their characteristics, in particular the systematic shortening of
their periods \citep{str04a,hrw+03,rhw+05,str05}, are difficult to
reconcile with them being AM CVn stars, but see
\citet{mn05}. Alternative models that have been proposed are a face-on
longer-period binary in which we only see the spin period of an
accreting white dwarf \citep{nhw02} and a pair of detached white
dwarfs generating X-ray emission through unipolar induction
\citep{wcr02}. The latter has recently attracted new attention, with
claims that the original model cannot explain the observations
\citep{mn05} as well as suggestions that in a somewhat different form
the model can at least explain RX J0806.3+1527 \citep{dis05}.

(Ultra-compact) X-ray binaries are rare objects. The known UCXBs are much
further away than the AM CVns stars which hinders optical studies. Therefore
it is only with the advent of 8m class telescopes that high-quality optical
spectra have become available, and consequently much of our current knowledge
is based on X-ray observations and photometry at different wavelengths. There
are 12 systems with known or suggested orbital periods. Four more systems are
identified based on similarities with known systems in either their X-ray
spectra \citep{jpc00} or as a result of their optical faintness in outburst,
as expected for the small accretion discs in these systems \citep{vm94}. Six
systems are found in a globular cluster \citep[see][for reviews of globular
cluster X-ray sources]{kdi+03,vl04}. 

\subsection{Why study ultra-compact binaries}

There are still many open questions in
the study of ultra-compact binaries which we would like to answer for the
following reasons: 
\begin{enumerate} 
\item Ultra-compact binaries are strong
gravitational-wave sources in the low-frequency regime where the ESA/NASA LISA
mission will be sensitive.
\item They are unique astrophysical laboratories because of the
combination of compact objects, short period variability and peculiar
chemical composition.
\item They are important test
of binary evolution theory, in particular the common-envelope phase.
\end{enumerate}

\begin{table*}
\caption{Overview of the UCXB (candidates). Data from \citet{vm94},
\citet{dma00}, \citet{lvv01}, \citet{kdi+03}, \citet{pou05} unless
otherwise indicated. }
\begin{tabular}{lllllllll}\hline
Name & Period & P/T & B & d & $M_V$ & \hspace*{-0.5cm}log $\frac{L_X}{L_{\rm Edd}}$ & remark & Ref. \\
 & (min) & & & (kpc) & & & & \\
\hline
4U 1820-30    & 11   & P & y & 7.6   & 3.7   & -0.7  & NGC 6624    & \\
4U 1543-624   & 18(?)& P & n & ?     & ?     & ?     & Ne-rich(?)  & 1 \\ 
4U 1850-087   & 21(?)& P & y & 8.2   & 5.1   & -2.4  & NGC 6712    & 2\\
M 15 X-2      & 23   & P & y & 10.3  & ?     & -2.2  & M 15        & 3, 4\\
XTE J1807-294 & 40   & T & n & ?     & ?     & ?     & msPSR         & \\
4U 1626-67    & 42   & P & n & $<4^a$& $>5.5$&$<-2.3$& \multicolumn{2}{l}{Ne-rich, slow PSR}\\
XTE J1751-305 & 42   & T & n & ?     & ?     & ?     & msPSR         &\\
XTE J0929-314 & 44   & T & n & ?     & ?     & ?     & msPSR         &\\
NGC 6652 B    & 44?  & P?& y?& 9.6   & 4.7   & -4.7  & MS colours  & 5\\
4U 1916-05    & 50   & P & y & 9     & 5     & -2    & Dipper      & \\ 
4U 0614+09    & 50?  & P & y & $<3$  & $>5.4$& $<-2$ & Ne-rich(?)  & 6, 7\\ 
XB 1832-330   &55?$^b$& P & y?& 9.6   & 3.7   & -2.3  & NGC 6652(A) & 5\\ \hline
XB 0512-401   & ?    & P & y & 12.1  &5.6$^c$& -2.0  & NGC 1851    & \\
2S 0918-549   & ?    & P & y & 4.2   & 6.9   & -2.5  & Ne-rich(?)  & 6\\
4U 1822-00    & ?    & P & n &$<20^d$&$>5.5$ & $<-1$ & Ne-rich(?)  & \\
XB 1905+000   & ?    & ? & y & 8     & 4.9   & -2.1  &             & 1\\ \hline
\end{tabular}
\textbf{Notes:} $a$ Assuming a distance from the plane $z < 1$kpc;
$b$ Or its aliases 2.2 and 4.4 hr; 
$c$ $M_B$; $d$ Assuming a reasonable maximum distance. For $d = 8$kpc, $M_V$ = 7.5 and log $L_X/L_{\rm Edd} = -1.7$\\
\textbf{References:} 1 \citet{wc04} and refs therein ; 2 \citet{hcn+96}; 3 \citet{dkz05}; 4 \citet{wa01}; 5 \citet{heg01}; 6 \citet{njm+04} and refs therein; 7 \citet{ojd+05}
\label{tab:candidates}
\end{table*}

\section{A short review on ultra-compact X-ray binaries}

\subsection{Observational properties}

An overview of the observational properties of the known and candidate
UCXBs is given in Table~\ref{tab:candidates}, ordered by (suggested)
orbital period (if known). The table lists the name, the orbital
period, the type of source (persistent versus transient), the
occurrence of type I X-ray bursts, the distance, absolute V-band
magnitude, X-ray luminosity in terms of the Eddington limit (taken as
$2 \times 10^{38}$ erg/s), remarks (msPSR = millisecond pulsar, for
discussion of the meaning of Ne-rich, see Sect.~\ref{X-ray}) and
references.

\subsubsection{X-ray properties}\label{X-ray}

UCXBs have so far been studied mainly in X-rays.  We can only
summarise the results briefly. There are two types of UCXBs:
persistent and transient systems. According to \citet{db03} systems
with periods longer than about 30 minutes should be transients
\emph{if the current mass-transfer rate is similar to its long-time
average}.  The observations show that there are at least 2 and
possibly 5 persistent systems with periods longer than 30
min. However, especially in comparing persistent and transient sources
there are many biases. Interestingly, the three transients all have
periods very close together between 40 and 44 min and all three of
them harbour accretion powered millisecond X--ray pulsars.

\textbf{(millisecond) pulsations}\\ An intriguing fact is that three
of the six known accreting millisecond pulsars reside in UCXBs, e.g.
\citet{mss+02,rss02}, while only 16 UCXBs are known. 4U 1626-67 is a
peculiar system because it harbours a young neutron star with a 7s
spin period.

\textbf{X-ray spectra}\\ The X-ray spectra lacked resolution until
recent observations obtained with the gratings on board the {\it
Chandra} satellite.  The X--ray spectra are difficult to interpret due
to the lack of clear emission features (except in the 7s accreting
pulsar 4U 1626-67 which shows strong O and Ne emission lines in its
X-ray spectrum \citep{sch+01}) and contributions of interstellar
absorption. A number of UCXBs have been (first) identified based on
their X-ray spectra, when \citet{jpc00} noted a similarity between the
spectrum of 4U 1850-087 and three more LMXBs.  The common feature in
the X-ray spectra was suggested to be due to an enhancement of neon in
these systems, however, this is uncertain. These observations were
interpreted as evidence for the donors being ONe, or more likely CO
white dwarfs in which Ne has sunk to the core in an earlier
evolutionary phase and is now exposed, after accretion has peeled--off
the outer layers \citep{ynh02}. Recently it has been noted that if it
is really the O/Ne ratio that is anomalous, and not the Ne abundance
itself, this points towards a He white dwarf donor, in which the O
abundance is strongly reduced due to CNO processing
\citep{icv05}. Interestingly, one system that shows a similar feature
in its X-ray spectrum \citep[4U~1556-605][]{ffm+03}, shows strong
hydrogen and helium emission in its optical spectrum, suggesting that
it is not an UCXB (Nelemans \& Jonker, in prep).

\textbf{Type I X-ray bursts}\\ For a number of UCXBs type I X-ray
bursts have been found, especially for the ones in globular clusters,
but that may be a selection effect. The peculiar chemical composition
of the accreting material will influence the burst properties and in
principle can give an extra tool to study the chemical composition
\citep[e.g.][]{cb01}. Recently a number of peculiar bursts have been
observed in 2S~0918-549 and 4U~0614+09 \citep{icv05,kuu05} which
\citet{icv05} suggest are due to thick layers of He that are
burned. It is clear that there is still uncertainty how we should
interpret the different pieces of information, because in some cases
the bursts and the optical spectra suggest conflicting compositions as
we will discuss below!

\subsubsection{Optical and UV properties}

\begin{figure*}
  \includegraphics[angle=-90,width=\textwidth,clip]{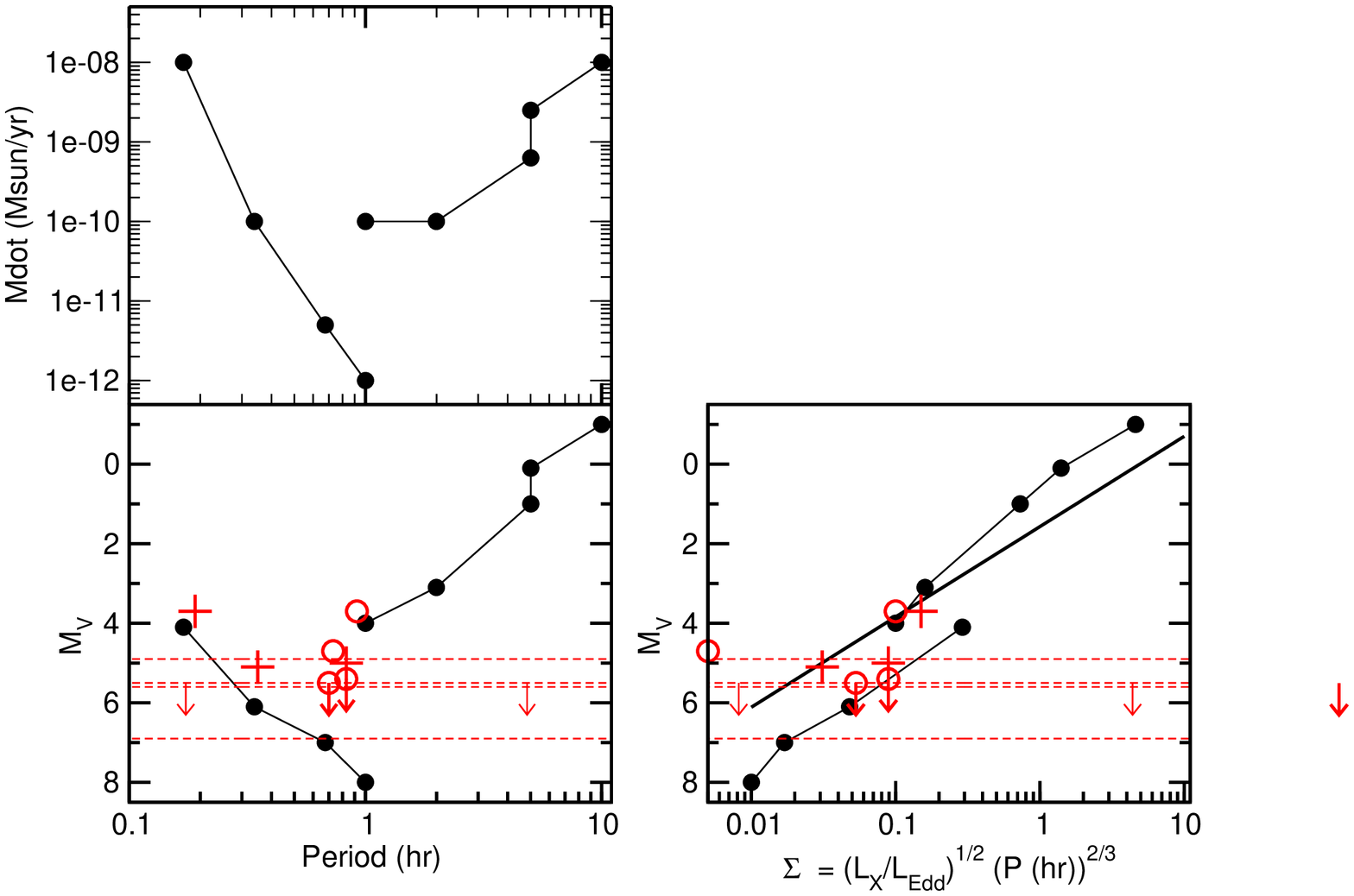}
  \caption{Absolute magnitudes as function of period (left) and
  $\Sigma$ (right) for a number of fiducial LMXBs and UCXBs with their
  mass transfer rate at the long-term average as expected from binary
  evolution calculations (solid dots). The observed (candidate) UCXBs
  are indicated with the plusses (period and magnitude reliable), open
  circles (periods and magnitudes not very reliable) and dashed line
  (periods unknown).  The straight line in the right panel gives the
  \citet{vm94} fit to the observations used by them.}
\label{fig:Mvs}
\end{figure*}

\textbf{Absolute magnitudes}\\ \citet{vm94} investigated the absolute
magnitudes of LMXBs, based on the assumption that most of the optical
flux is generated by X-ray irradiation of the disc. This means that
the two dominant factors determining the absolute magnitude of LMXBs
are the X-ray luminosity and the size of the disc. They derived a
parameter $\Sigma \propto L_X^{1/2} P_{\rm orb}^{2/3}$ which should
correlate with the absolute magnitude. They indeed found a strong
correlation.  Because of their small disc size UCXBs are expected to
be relatively dim. This has served as an indication for systems to be
classified as UCXB candidates. Based on the irradiation model of
\citet{ak93} we calculated the absolute magnitudes of a number of
fiducial LMXBs which we compare to the observed systems
(Fig.~\ref{fig:Mvs}).  Indeed most sources seem to be consistent with
being UCXBs, although NGC 6652 B (see Sect.~\ref{GC}) is such a faint
X-ray source that the irradiated disc model is probably not applicable
for this source and the system more likely is a quiescent LMXB
\citep[see][]{heg01}.

\textbf{Photometry}\\ The faintness of the optical counterparts has inhibited
high-quality spectra for a long time, but photometric studies have revealed
many of the know orbital periods. Recent developments in photometry are the
used of ULTRACAM to study 4U~0614+09, yielding a suggested orbital periods of
50 min \citep{ojd+05}. Recent photometry of 4U~1543-624 yielded a possible
orbital period of 18 min \citep{wc04}.

\begin{figure*}
  \includegraphics[angle=-90,width=\textwidth]{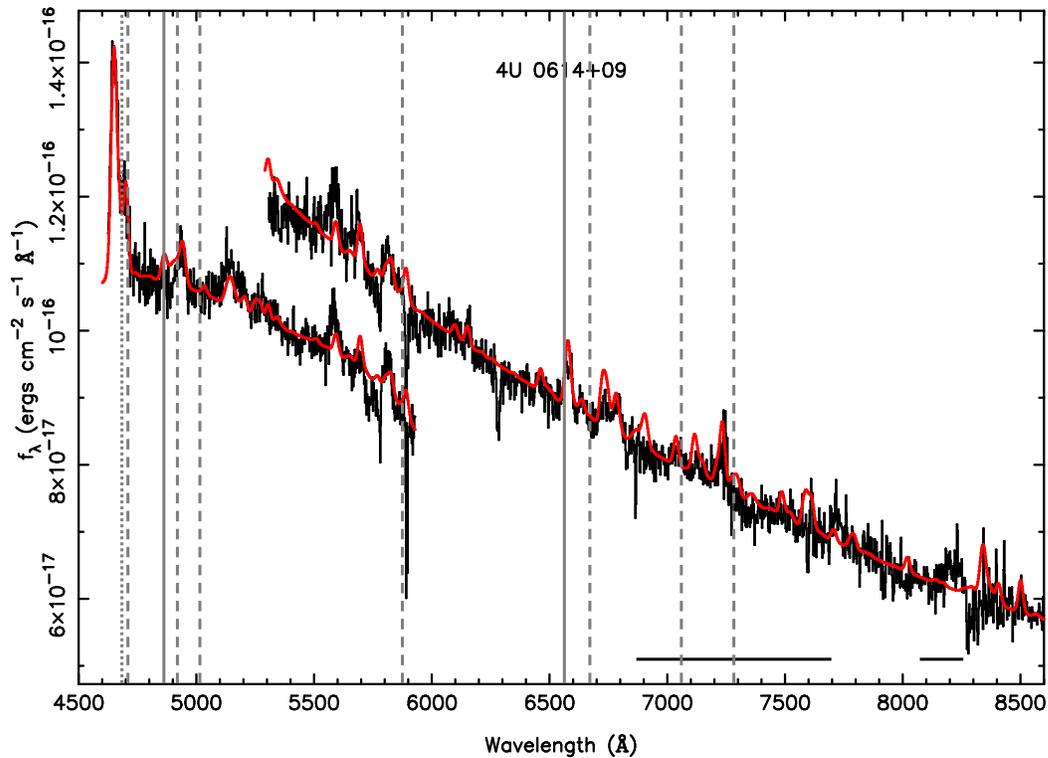}
  \caption{VLT spectra of 4U 0614+09, showing lines from a
  carbon-oxygen accretion disc. From \citet{njm+04}}
\label{fig:2003}
\end{figure*}

\textbf{Spectra}\\ We recently started a systematic spectroscopic
study of (candidate) UCXBs. The main aim was to confirm/reject
candidates and to study the chemical composition of the donors stars
in these systems, which hold clues to the formation of UCXBs (see
Sect.~\ref{formation}). The first results are published in
\citet{njm+04} and are summarised in Fig.~\ref{fig:2003}. We
identified the features in the spectrum of 4U 0614+09 as relatively
low ionisation states of carbon and oxygen. This clearly identifies
this system as an UCXB and suggests the donor in this system is a
carbon-oxygen white dwarf. The similarity of the spectrum of 4U
1543-624 suggests it is a similar system, while for 2S 0918-549 the
spectrum didn't have a high enough S/N ratio to draw firm conclusions,
but it is also is consistent with being a similar system (and clearly
does not show the characteristic strong hydrogen emission lines of
low-mass X-ray binaries). We therefore concluded that all these
systems are UCXBs. In Sect.~\ref{results} we present new results.

\subsubsection{Globular cluster sources}\label{GC}

A special mention should be made about UCXBs in globular clusters
(GCs). Five of the 12 systems with known orbital periods reside in
GCs, an even larger enhancement compared to the disc than for
longer-period X-ray binaries. For reviews of X-ray sources in globular
clusters see \citet{ver04,vl04}. The shortest period system
4U~1820-30, which with an orbital period of 11 min for a long time has
been the shortest period binary known and is the most well-known
source of this class. The positional accuracy and better sensitivity
of {\it Chandra} and {\it XMM} are clearly crucial for better
understanding these systems. For instance in M 15 the bright X-ray
source has been resolved in two \citep{wa01}, the brightest one
associated with a faint FUV source that shows a clear modulation on 23
min \citep{dkz05}.  In NGC 6652 the suggested counterpart of the
brightest source turns out to be the counterpart of a (much fainter)
X-ray source \citep[source B][]{heg01}. It has a suggested periodicity
of 44 min \citep{dma00} but its colours are consistent with a
main-sequence star \citep{heg01}. The counterpart of the bright source
(source A), has a possible period of 55 min \citep[but with aliases at
2.2 and 4.4 hr][]{heg01}.

\subsection{Formation and evolution}\label{formation}

For a detailed discussion of the formation of AM~CVn systems and UCXBs
in the field we refer to \citet{npv+00,prp02} and references
therein. In short there are three routes, differentiated by the nature
of the donor star: (i) a white dwarf donor when a detached binary with
a white dwarf and either a neutron star/black hole or white dwarf
comes into contact due to angular momentum losses via gravitational-wave radiation; (ii) a (semi-degenerate) helium star donor that
evolved from a helium core burning star that filled its Roche lobe to
a white dwarf or neutron star/black hole and (iii) the core of a star
that filled its Roche lobe to a white dwarf or neutron star/black hole
at the end of the main sequence and thus has a helium rich core. In
globular clusters the formation of X-ray binaries is probably
dominated by dynamical interactions \citep[see][]{ver04}.

These different formation scenarios result in principle in different
chemical compositions, but there is some overlap. The three formation
scenarios will yield the following chemical composition:
\begin{enumerate}
\item White dwarf donors: He + CNO processed (i.e. mainly N) or C/O
\item He star donor: He + He burning products
\item Evolved secondaries: He + CNO processed + some H
\end{enumerate}

\subsection{Open questions}

The rapid increase in the number of (candidate) UCXBs in recent years
shows that this is an area where new technological developments play
an important role. However, with the increase in the sample size a
number of new questions have risen, while many of the old questions
still remain. A subjective list of the current most important open
questions is
\begin{itemize}
\item What are the chemical compositions of the donor stars and what
  are the best measurements to determine this? In particular the
  interpretation of the optical spectra (see below) and the type I
  X-ray bursts deserve attention.
\item What are the orbital periods of the remaining systems? Only when
  we have found these periods we can really start to look at the
  period distribution. The peculiar fact that most systems had an
  orbital period of about 40 min is already being diluted by new
  period determinations at 18 and 23 min. However, still the
  clustering of the three transients at very similar periods remains.
\item What is the determining factor in the difference between
  transient and persistent sources?
\item Why are so many of the accreting millisecond pulsars found in
UCXBs? Is this due to the low average mass accretion rate that is
expected in the longest-period UCXBs?
\end{itemize}

\section{VLT spectroscopy of ultra-compact X-ray binaries}\label{results}

In order to tackle some of the open questions mentioned in the last
section, we continued our spectroscopic study of (candidate) UCXBs.

\subsection{Observations}

We used the FORS2 spectrograph on the 8.2m Very Large Telescope (VLT)
of the European Southern Observatory at Paranal to obtain optical
spectra of our candidate UCXBs. The observations were taken in the
spring of 2003 and 2004. 

\subsection{Results}

\begin{figure*}
\begin{center}
  \includegraphics[angle=-90,width=0.9\textwidth]{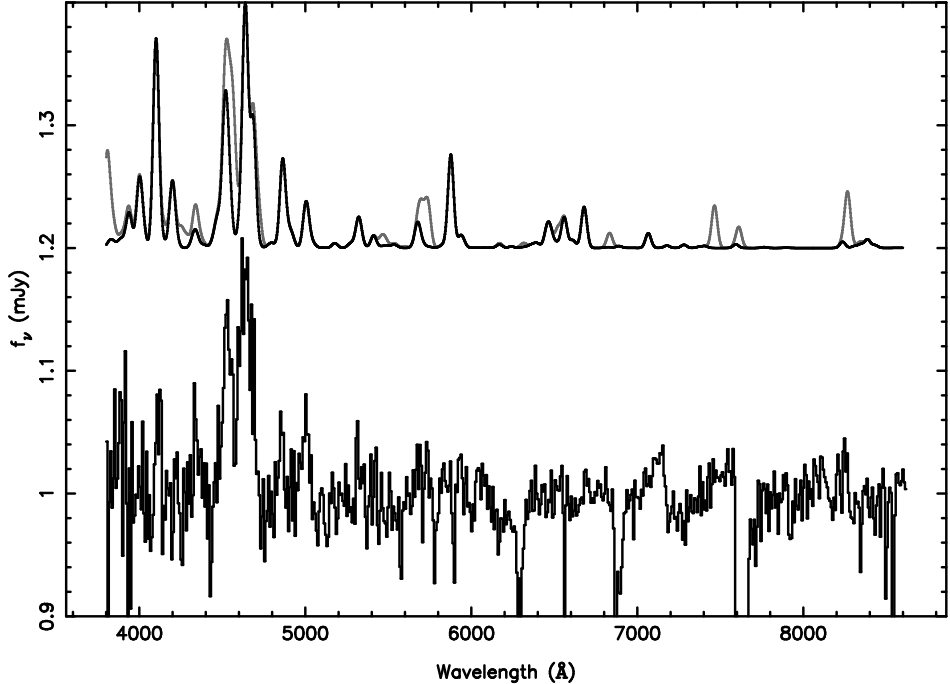}
  \caption{VLT spectra of 4U 1916-05 (bottom). Above we show an LTE
  model with only helium and nitrogen (black line) and a similar model
  with all metals at solar abundance (grey line).}
\label{fig:1916}
\end{center}
\end{figure*}

\textbf{4U 1916-05}\\ In 2004 we obtained a spectrum of the known UCXB
4U 1916-05, which has a 50 min orbital period. In Fig.~\ref{fig:1916}
we show the spectrum together with an LTE model of a disc of pure
helium with (overabundant) nitrogen (black line) and a similar model
with metals at solar abundance (grey line). We conclude that this is
the first clear detection of a helium donor in an optical
spectrum. However, the spectrum does not look at all like we had
anticipated (i.e. strong helium emission lines, like the spectra of AM
CVn stars ES Cet and GP Com). This means that really high S/N spectra
are needed to determine the chemical composition of the donors and
that in our earlier results, the spectrum of 2S 0918-549 clearly is
not good enough to distinguish between helium and C/O and might be
more likely a helium donor because of the properties of its X-ray
bursts \citep{icv05}.

\begin{figure*}
\begin{center}
  \includegraphics[angle=-90,width=0.8\textwidth]{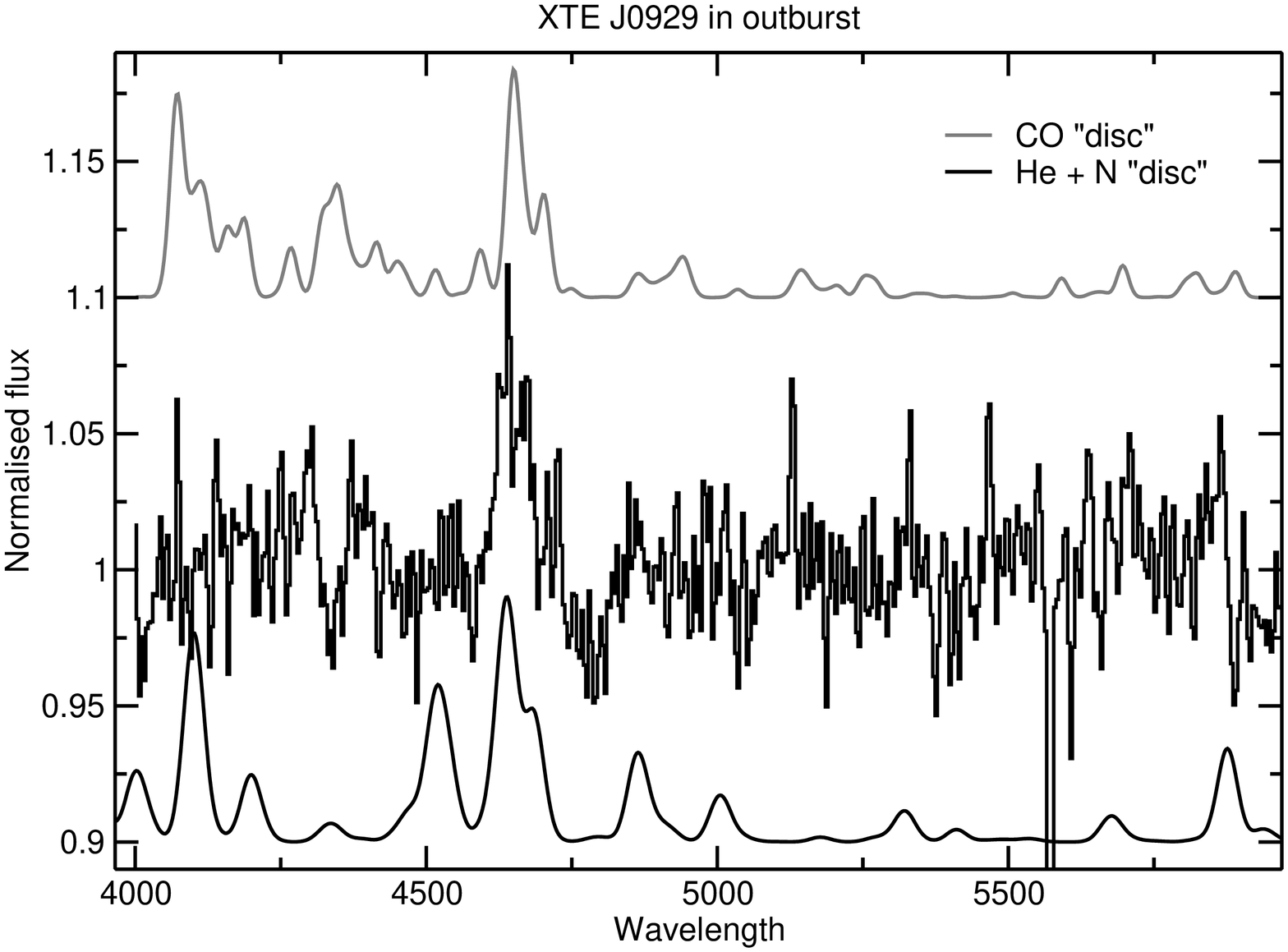}
  \caption{Spectrum of XTE J0929-314 taken during outburst. Above and
  below an LTE model for a CO (top) and a He/N (bottom) disc are
  shown.}
\label{fig:0929}
\end{center}
\end{figure*}

\begin{figure}
\begin{center}
  \includegraphics[angle=-90,width=\columnwidth,clip]{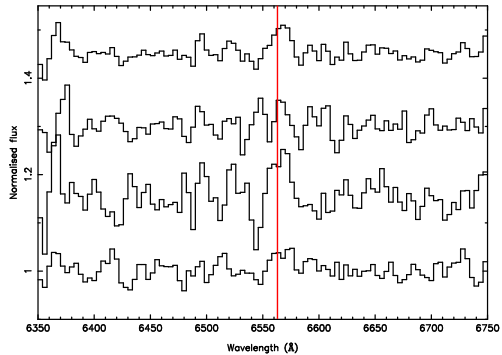}
  \caption{Region around H$\alpha$ of the outburst spectrum of XTE
  J0929-314 on the basis of which \citet{ccg+02} claim H$\alpha$
  emission which would contradict the interpretation of the donors as
  white dwarfs. The 3 individual spectra are shown at the bottom, the
  average as the top spectrum.}
\label{fig:0929_Ha}
\end{center}
\end{figure}

\textbf{XTE J0929-314}\\ XTE J0929-314 is one of the three transient UCXBs
and was discovered in 2002 \citep{rss02}. In outburst a spectrum was
taken with the ESO 3.6m telescope in which emission at H$\alpha$ and
around 4650\AA~was reported \citep{ccg+02}. In particular the
H$\alpha$ emission is important, since this would clearly point to a
donor with still some hydrogen left. However, the evidence for
H$\alpha$ emission is rather weak as is shown in
Fig.~\ref{fig:0929_Ha}, where we display the spectra. The formal
significance of the H$\alpha$ emission is 3$\sigma$, but one has to
keep in mind that there are many emission lines in the night sky
around H$\alpha$ that could lead to systematic errors. The emission
around 4650\AA~is clearly significant, as seen in Fig.~\ref{fig:0929},
where we also plot simple C/O and He LTE model. The spectrum is not
good enough to determine the composition, although the lack of
emission around 4500\AA~suggest a C/O donor.

\begin{figure*}
 \begin{center}
  \includegraphics[angle=-90,width=0.8\textwidth]{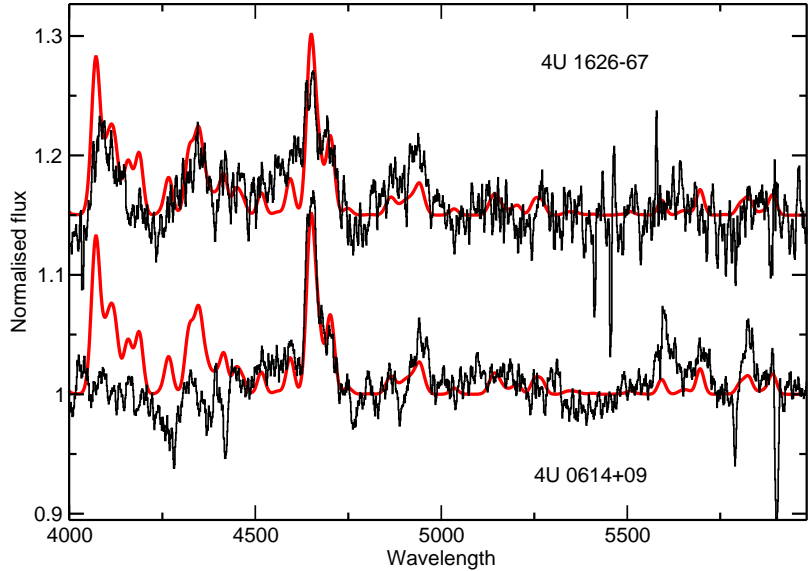}
  
  \caption{Spectra of 4U 1626-67 and 4U 0614+09. The spectra are quite
  similar above 4500\AA, but below they are very different and
  interestingly the spectrum of 4U 1626-67 follows the LTE model that
  we used for the line identification of 4U 0614+09 much better than
  4U 0614+09 itself!}
\label{fig:1626_0614}
 \end{center}
\end{figure*}

\textbf{4U 1626-67 and 4U 0614+09}\\ We also obtained a spectrum of 4U
1626-67. In \citet{nj04} we already showed a comparison between the
spectrum of 4U 1626-67 and our earlier 4U~0614+09 spectrum which
looked very much like each other, in agreement with the interpretation
of the donors of both these systems being C/O rich. Indeed 4U 1626-67
follows the simple C/O LTE model that we used for 4U 0614+09
reasonably well even at bluer wavelengths than out 4U 0614+09 spectrum
reached. However, a second spectrum of 4U 0614+09 (obtained with the
VLT by Klaus Werner, Werner et al.~in prep.) with the same setup and
wavelength coverage as our 4U 1626-67 spectrum clearly does
\emph{not} follow the simple LTE model! (see
Fig.~\ref{fig:1626_0614}). It is clear that the last word on the
interpretation of these spectra is not yet said.

\bibliography{journals,binaries} 
\bibliographystyle{apj}

\end{document}